\documentclass[prb,amsmath,amssymb,showpacs,preprint]{revtex4}

\usepackage{graphicx}
\def\be{\begin{equation}}
\def\ee{\end{equation}}
\def\ni{\noindent}
\def\sf{{\tt *)}\newline}
\def\ep{\epsilon}

\begin{document}
\title{Derivation of the Langevin equation from the principle of detailed balance}
\author{Jorge Berger}
\affiliation{Department of Physics and Optical Engineering, Ort Braude College, P. O. Box 78,
21982 Karmiel, Israel}
\email{jorge.berger@braude.ac.il}
\begin{abstract}
For a system at given temperature, with energy known as a function of a set of variables, we obtain the thermal fluctuation of the evolution of the variables by replacing the phase-space with a lattice and invoking the principle of detailed balance.
Besides its simplicity, the asset of this method is that it enables us to obtain the Langevin equation when the phase-space is anisotropic and when the system is described by means of curvilinear coordinates. As an illustration, we apply our results to the Kramer--Watts-Tobin equation in superconductivity. The choice between the It\^{o} and the Stratonovich procedures is discussed.
\end{abstract}
\maketitle

\section{INTRODUCTION}
In this article we deal with what we call ``purely diffusive systems" in contact with a thermal bath. By this we mean systems with a state described by microscopic variables $x_1,\dots x_N$, with energy $E(x_1,\dots x_N)$, which in the absence of thermal fluctuations are expected to follow an evolution equation
\be
\frac{dx_j}{dt}=-\Gamma _j(x_1,\dots x_N)\frac{\partial E}{\partial x_j} \;,
\label{dissip}
\ee
where $t$ is the time and the positive coefficients $\Gamma_j$ are determined by the dynamics of the system and its interaction with its environment.
$\partial E/\partial x_j$ may be regarded as a driving force and $\Gamma_j$ as a compliance coefficient.
The Langevin approach tells us that the influence of thermal fluctuations can be taken into account by adding a fluctuating quantity at the right hand side of the evolution equation; this fluctuating quantity is called the ``Langevin term."

The paradigm of a purely diffusive system is a particle that undergoes Brownian motion. In this case the variable is its momentum, $E$ is its kinetic energy and $\Gamma$ is the Stokes coefficient. Paul Langevin dealt with this problem\cite{Lang} and determined the variance of the Langevin term by invoking the theorem of the equipartition of the kinetic energy among the various degrees of freedom of a system in thermal equilibrium. Gillespie\cite{Gi} notes that it is not obvious that the influence of fluctuations can be separated as an additive term with zero average; regarding the velocity evolution as a Markovian process and assuming that the ``stepping functions" (will be defined in the following section) are linear functions of velocity, it is shown that this separation indeed occurs.
Katayama and Terauti\cite{KT} used the Langevin equation to study Brownian motion of a single particle under steady plane shear flow. Balescu\cite{Bal} introduces a Langevin equation in a model for the description of a plasma. Bringuier\cite{BrinI} discusses the difficulties encountered when applying the Langevin approach to the Hall effect. The relation between the Langevin and the Klein--Kramers approaches is discussed in Ref.~\onlinecite{BrinII}.

During the century that has elapsed since Langevin's paper, his approach has been extended to wide classes of problems in Physics\cite{Chand,HH,Cof,relat} and the Langevin term is determined by means of the fluctuation-dissipation theorem.\cite{FDT1,FDT2,FDT3} In this paper we will evaluate the distribution of the Langevin term by means of the principle of detailed balance.\cite{FDT2}

The tools for handling the problems considered in this paper may be found in the literature on stochastic differential equations (e.g. Refs. \onlinecite{Str,Schu,Oks}) and many of the questions raised here may be avoided by switching to the Fokker–-Planck equation; this article is addressed to
those readers, pressumably physicists, who prefer a more intuitive approach.

\section{OUR METHOD}
\subsection{1D case\label{sec1D}}
We consider a one-dimensional system with microscopic state determined by the variable $x$. We discretize $x$ and assume that its possible values are $x_i=i\ell$, where $\ell$ is a ``lattice constant." We denote by $\ep_i=E(x_{i+1})-E(x_i)$ the energy difference between consecutive lattice points. We assume now that for a short period of time $\delta t$ the probability of passing from $x_i$ to $x_{i\pm 1}$ is given by $W_{i\pm}\delta t$, where $W_{i+}$ and $W_{i-}$ are ``stepping functions" that stand for the transition probability rates.

The principle of detailed balance asserts that in thermal equilibrium the probability for a transition from $i$ to $i+1$ equals that for a transition in the opposite direction, i.e., denoting by $P^{\rm eq}_i$ the equilibrium probability for the value $x=x_i$, $P^{\rm eq}_iW_{i,+}=P^{\rm eq}_{i+1}W_{i+1,-}$. Since $P^{\rm eq}_i/P^{\rm eq}_{i+1}=e^{\ep_i/k_BT}$, where $k_B$ is the Boltzmann constant and $T$ is the temperature,
\be
W_{i+1,-}=W_{i,+}e^{\ep_i/k_BT} \;.
\label{i}
\ee

It should be noted that not every system obeys detailed balance. Denoting by $P_i$ the probability for $x=x_i$ (not necessarily for equilibrium), $P_iW_{i,+}-P_{i+1}W_{i+1,-}$ stands for the probability current. Detailed balance requires that this current vanishes, whereas in order to mantain a stationary state it is sufficient that the divergence of the current vanishes. If probability currents are present in equilibrium, it follows that there are driving forces which cannot be expressed as the gradient of the energy as in Eq.~(\ref{dissip}) (such as the magnetic force on a charged particle). Therefore, systems that do not obey detailed balance are beyond the scope of this article.

In order to obtain more symmetric expressions, we write $W_{i,+}=w_i \lambda_i$, $W_{i,-}=w_i/\lambda_i$ and, taking $\ell$ sufficiently small so that quantities of order $O(\ell^2)$ can be dropped, we write $w_{i\pm 1}=w_i\pm w'_i$ and $\lambda_{i\pm 1}=\lambda_i\pm \lambda '_i$, where $w'_i/w_i$ and $\lambda '_i$ are at most of order $O(\ell)$. With this notation and approximation, Eq.~(\ref{i}) becomes
\be
\frac{w_i+w'_i}{\lambda_i+ \lambda '_i}=w_i \lambda_i e^{\ep_i/k_BT} \;.
\label{ip1}
\ee
Similarly, requiring detailed balance between the sites $i$ and $i-1$ we obtain
\be
(w_i-w'_i)(\lambda_i-\lambda '_i) e^{\ep_i/k_BT}=\frac{w_i}{ \lambda_i} \;,
\label{im1}
\ee
where we have exchanged sides in the equation and neglected the $O(\ell^2)$ difference $\ep_{i-1}-\ep_i$. Multiplying the equations (\ref{ip1}) and (\ref{im1}) and neglecting the $O(\ell^2)$ term $(w'_i/w_i)^2$ we obtain  
\be
\frac{\lambda_i-\lambda '_i}{\lambda_i+ \lambda '_i}=1 \;,
\ee
hence $\lambda '_i$ is of order $O(\ell^2)$ and will be dropped. $\lambda_i$ can now be obtained from Eq.~(\ref{ip1}); keeping terms of order $O(\ell)$ and making use of the definitions of $w'_i$ and $\ep_i$, it becomes 
\begin{eqnarray}
\lambda_i&=&\sqrt{(1+w'_i/w_i)e^{-\ep_i/k_BT}}=1+w'_i/(2w_i)-\ep_i/(2k_BT) \nonumber \\
&=&1+(\ell/2) d[\log (w)]/dx-[\ell/(2k_BT)]dE/dx \;,
\end{eqnarray}
where we have defined a smooth function $w$ such that $w(x_i)=w_i$.

Let us denote by $\delta x$ the increment of the variable $x$ during the period of time $\delta t$. For sufficiently small $\delta t$ we can neglect multiple transitions and the possible values of $\delta x$ are 0 and $\pm\ell$. The average value of $\delta x$ will be
\be
\langle \delta x \rangle =\ell w_i \delta t(\lambda_i-1/\lambda_i)=\ell^2 w \delta t[d(\log w)/dx-(1/(k_BT))dE/dx] \;,
\label{av1}
\ee
where in the last step we have neglected higher orders of $\ell$ and have dropped the index $i$. Similarly, the variance of $\delta x$ will be
\be
\langle (\delta x)^2 \rangle =\ell^2 w_i \delta t(\lambda_i+1/\lambda_i)=2\ell^2 w \delta t\;,
\label{var1}
\ee
where besides dropping terms that are of higher order in $\ell$ we have used the fact that, for small $\delta t$, $\langle \delta x \rangle^2\ll\langle (\delta x)^2 \rangle$.

We now get rid of the unphysical lattice by defining $\Gamma (x)=\ell^2 w(x)/(k_BT)$. With this notation Eqs. (\ref{av1}) and (\ref{var1}) become
\be
\langle \delta x \rangle =\Gamma (x) \delta t[k_BT d(\log \Gamma)/dx-dE/dx] 
\label{av1g}
\ee
and
\be
\langle (\delta x)^2 \rangle =2k_BT \Gamma (x) \delta t \;.
\label{var1g}
\ee
Finally, we consider a lapse of time $\tau$ which is very short compared with the relaxation time, but very long compared with $\delta t$. By the central limit theorem,\cite{FDT2,CL,Fel} the increment of $x$ [which is the sum of many increments described by Eq.~(\ref{av1g})] will be
\be
\Delta x =\Gamma (x) [k_BT d(\log \Gamma)/dx-dE/dx] \tau +\eta \;,
\label{AV1}
\ee
where $\eta$ is a fluctuating term with average 0, variance $2k_BT \Gamma (x)\tau$ and Gaussian distribution. $\eta$ is the Langevin distribution we were looking for. There is still a subtle question concerning the precise value of $x$ at which $\Gamma (x)$ has to be evaluated; this issue is considered in Appendix \ref{Ito}.

Let us now compare the nonfluctuating part of Eq.~(\ref{AV1}) with Eq.~(\ref{dissip}). If $\Gamma$ is independent of $x$, Eq.~(\ref{dissip}) is recovered; otherwise, there is also a drift term $k_BT d(\log \Gamma)/dx$ that pushes $x$ towards values where $\Gamma$ is larger. The drift term can be absorbed into Eq.~(\ref{dissip}) if we replace the energy $E$ by $G=E-k_BT\log \Gamma$. The term $k_BT\log \Gamma$ may be regarded as a sort of chemical potential, where $\Gamma$ plays the role of the activity.

\subsection{Multivariable system}
We consider now a system with variables $x_1,\dots x_N$. For each of the variables we can repeat the analysis of the previous section and Eq.~(\ref{AV1}) generalizes to
\be
\Delta x_j =-\Gamma_j (\partial G_j/\partial x_j) \tau +\eta_j \;,
\label{AVN}
\ee
with $G_j=E-k_BT\log \Gamma_j$ and $\eta_j$ is a Langevin function with average 0, variance $2k_BT \Gamma_j\tau$ and Gaussian distribution.

We might also be interested in the evolution of other variables rather than those in the set $x_1,\dots x_N$. This problem is considered in Appendix \ref{transf}.

\subsection{Curvilinear coordinates}
Now the volume in phase space is proportional to the Jacobian of the coordinates; therefore, different lattice points in the discretized phase space may represent different volumes. As a consequence, $P^{\rm eq}_i$ has to be multiplied by this Jacobian and equations like Eq.~(\ref{i}) have to be modified accordingly. Before we deal with the general case, let us consider the case of polar coordinates.
\subsubsection{Polar coordinates\label{polar}}
Let the coordinates be $r,\varphi $, with volume element $rdrd\varphi $. For the variable $\varphi $ the analysis remains unchanged, but for $r$ we have $P^{\rm eq}_i/P^{\rm eq}_{i+1}=e^{\ep_i/k_BT}r/(r+\ell)$ and equations like Eq.~(\ref{i}) have to be replaced with equations like
\be
W_{i+1,-}=W_{i,+}e^{\ep_i/k_BT}r/(r+\ell) \;.
\label{ipol}
\ee
Following the steps of Sec.~\ref{sec1D}, instead of Eq.~(\ref{AV1}) we now obtain
\be
\Delta r =\Gamma_r \partial[k_BT \log (r\Gamma_r) - E]/\partial r\, \tau +\eta_r \;,
\label{AV1pol}
\ee
where $\eta_r$ has average 0, variance $2k_BT \Gamma_r\tau$ and Gaussian distribution. The replacement of the term $\log \Gamma_r$ with $\log r\Gamma_r$ implies a drift towards larger values of $r$.

It is tempting\cite{Engel} to attribute the drift towards larger values of $r$ to the fluctuations of $\varphi $: if the system moves in phase space by the amount $r\delta \varphi $ perpendicular to the radial direction, the new value of $r$ would be $\sqrt{r^2+(r\delta \varphi)^2}\approx r[1+(\delta \varphi)^2/2]$. After a lapse of time $\tau$ this effect would contribute an increment of $r$ by the amount $r\langle (\Delta \varphi)^2/2\rangle=rk_BT \Gamma_\varphi\tau$. Comparison of this result with Eq.~(\ref{AV1pol}) indicates that this interpretation would be consistent with the principle of detailed balance only if $\Gamma_r=r^2\Gamma_\varphi$.

\subsubsection{General case}
Let the coordinates be $v_1,\dots v_N$, with volume element $J(v_1,\dots v_N)dv_1,\dots dv_N$. Then, when dealing with the transitions of the variable $v_j$, Eq.~(\ref{ipol}) generalizes to
\be
W_{i+1,-}=W_{i,+}e^{\ep_i/k_BT}J/(J+\ell\partial J/\partial v_j) \;.
\label{igen}
\ee
Following the steps of Sec.~\ref{sec1D}, $\lambda _i$ has an additional term $[\ell/(2J)]\partial J/\partial v_j$ and Eq.~(\ref{AV1pol}) generalizes to
\be
\Delta v_j =\Gamma_j \partial[k_BT \log (J\Gamma_j)- E]/\partial v_j\, \tau +\eta_j \;,
\label{AV1gen}
\ee
where $\eta_j$ has average 0, variance $2k_BT \Gamma_j\tau$ and Gaussian distribution.
An analogous result for macroscopic variables was obtained in Ref.~\onlinecite{GGG}.

A Mathematica-program that illustrates the use of this result is provided in Appendix~\ref{test}.

\section{Application---A model for superconductivity}
One of the most useful models in the study of dynamic properties of superconductors is the time-dependent Ginzburg--Landau model.\cite{GL,Tinkham} In this model the microstate of a superconductor is described by a complex field $\psi (x,y,z)$, such that $|\psi (x,y,z)|^2$ is proportional to the density of superconducting electrons at position $(x,y,z)$. Knowledge of the field $\psi$ enables us to evaluate several measurable quantities, such as the supercurrent density. Since in this model the variable of the problem is itself a field, it may provide an example in which the Langevin approach appears to be more practical than the Fokker–-Planck equation, since the latter is a partial differential equation in a space with infinitely many dimensions.

In most cases, the time-dependent Ginzburg--Landau model is justified only for temperatures very close to the transition temperature. The model was generalized by Kramer and Watts-Tobin;\cite{Kramer} this generalized model is expected to be valid as long as there is local equilibrium.
In order to focus on the aspects that we want to illustrate, we deal here with a simplified situation of the Kramer--Watts-Tobin model. We consider a uniform 1D superconductor with periodic boundary conditions and ignore the electromagnetic field. We discretize the system by dividing it into $N$ segments of equal length and denote by $\psi_j$ the value of $\psi $ at segment $j$. With appropriate normalizations, the energy of the system is given by
\be
E=\sum_{j=1}^N (-|\psi_j|^2+\frac{1}{2}|\psi_j|^4+\xi^2|\psi_j-\psi_{j-1}|^2) \;,
\ee
where $\xi$ is a constant that depends on the material, the temperature, and the length of each segment. If fluctuations are ignored, the evolution of $\psi_j$ is given by\cite{Kramer}
\be
\frac{u}{\sqrt{1+\gamma ^2|\psi_j|^2}}\left(\frac{d\psi_j}{dt}+\frac{\gamma ^2}{2}\frac{d|\psi_j|^2}{dt}\psi_j\right)=(1-|\psi_j|^2)\psi_j+\xi^2(\psi_{j+1}+\psi_{j-1}-2\psi_j) \;,
\label{KWT}
\ee
where $u$ and $\gamma $ are additional positive constants of the model.

Let us first consider the case $\gamma ^2\langle |\psi_j|^2\rangle\ll 1$, so that the left hand side in Eq.~(\ref{KWT}) can be approximated by $u d\psi_j/dt$. In this case it is convenient to express $\psi_j$ in Cartesian form, $\psi_j=x_j+i y_j$ and the energy becomes $\sum_{j=1}^N [-(x_j^2+y_j^2)+(x_j^2+y_j^2)^2/2+\xi^2(x_j-x_{j-1})^2+\xi^2(y_j-y_{j-1})^2]$. Performing the derivatives and separating real and imaginary parts, Eq.~(\ref{KWT}) takes the form
\be
\frac{dx_j}{dt}=-\Gamma_x\frac{\partial E}{\partial x_j} \;,
\label{KW0}
\ee
with $\Gamma_x=1/(2u)$, and an analogous equation is obtained for $y_j$. Since $u$ is constant, we can apply the result (\ref{AV1}) with no drift term and conclude that after time $\tau$ the fluctuating part of the increment of $x_j$ will have a variance $k_BT\tau/u$.

Let us now consider the general situation. In this case it is convenient to express $\psi_j$ in polar form, $\psi_j=r_j e^{i\varphi _j}$, the energy becomes $\sum_{j=1}^N [-r_j^2+r_j^4/2+\xi^2(r_j^2+r_{j-1}^2-2r_j r_{j-1}\cos(\varphi_j-\varphi_{j-1}))]$ and the expression in brackets at the left hand side of Eq.~(\ref{KWT}) becomes $[(dr_j/dt)(1+\gamma ^2 r_j^2)+ir_jd\varphi _j/dt ]e^{i\varphi _j}$. Multiplying Eq.~(\ref{KWT}) by $e^{-i\varphi _j}$ and taking the real part we obtain the evolution of $r_j$,
\be
\frac{dr_j}{dt}=-\frac{1}{2u\sqrt{1+\gamma^2 r_j^2}}\frac{\partial E}{\partial r_j} \;;
\label{KWr}
\ee
taking the imaginary part gives the evolution of $\varphi _j$,
\be
\frac{d\varphi _j}{dt}=-\frac{\sqrt{1+\gamma^2 r_j^2}}{2u r_j^2}\frac{\partial E}{\partial \varphi _j} \;.
\label{KWphi}
\ee
These equations are in the form of Eq.~(\ref{dissip}), with $\Gamma _r=1/(2u\sqrt{1+\gamma^2 r_j^2})$ and $\Gamma _\varphi =\sqrt{1+\gamma^2 r_j^2}/(2u r_j^2)$. In the extreme case $\gamma r_j\gg 1$,
$\Gamma _r\approx 1/(2u\gamma r_j)$ and $\Gamma _\varphi\approx \gamma/(2u r_j)$. Since in this limiting situation $r_j \Gamma _r$ does not depend on $r_j$ and $r_j \Gamma _\varphi $ does not depend on $\varphi _j$, the correction terms $\log(r \Gamma )$ are not required and the formalism developed in Sec.~\ref{polar} can be applied with no drift terms. For general $\gamma $, thermal fluctuations add a drift to Eq.~(\ref{KWr}) and lead to
\be
\Delta r_j =\Gamma_r \left[k_BT \left(\frac{1}{r_j}-\frac{\gamma ^2r_j}{1+\gamma^2 r_j^2}\right) -\frac{\partial E}{\partial r_j}\right] \tau +\eta_r 
=\Gamma_r \left[\frac{k_BT}{r_j(1+\gamma^2 r_j^2)} -\frac{\partial E}{\partial r_j}\right] \tau +\eta_r   \;,
\label{KWrgen}
\ee
where $\eta_r$ is the usual Langevin term with variance $2k_BT \Gamma_r\tau$.

It should be emphasized that Eq.~(\ref{KWrgen}) (including the drift) is an extension of Eq.~(\ref{KWr}) and not a modification of it. As an illustration of this statement, let us focus on the case $\gamma =0$ already considered in Eq.~(\ref{KW0}). In this case the drift term becomes $k_BT/r_j\neq 0$ and $\Gamma_r$ becomes $1/(2u)$, i.e. $\Gamma_r=\Gamma_x$. Moreover, let us for a moment leave the KWT model aside and consider the toy model $E= \sum_{j=1}^N r_j^2=\sum_{j=1}^N (x_j^2+y_j^2)$, i.e., we just have $ud\psi_j/dt= -\psi_j$ instead of Eq.~(\ref{KWT}) and $udr_j/dt= -r_j$ instead of Eq.~(\ref{KWr}). It follows that if the drift term $k_BT/r_j$ were not present, the evolution equations for $x_j$, $y_j$ and $r_j$ (including fluctuations) would all become identical and we would therefore have $\langle r_j^2\rangle=\langle x_j^2\rangle=\langle y_j^2\rangle$, whereas the true relationship is $\langle r_j^2\rangle=\langle x_j^2+y_j^2\rangle=2\langle x_j^2\rangle$.

\section{Conclusion}
We have developed a simple method that enables us to derive the Langevin and drift terms for systems with random-walk type evolution. The cocepts of probability theory that we have invoked are intuitive, elementary and in the ``language" used by undergraduate Physics textbooks.
This method is particularly useful when the dynamics leads to an anisotropic phase space or when the evolution is naturally expessed in curvilinear coordinates.

\begin{acknowledgments}
I have benefited from correspondence with Eric Bringuier, Moshe Gitterman, Peter H\"{a}nggi, Grzegorz Jung, Eduardo Mayer-Wolf, Zeev Schuss and Roman Vorobyov.
\end{acknowledgments}

\appendix
\section{It\^{o} or Stratonovich?\label{Ito}}
We first note that both terms in Eq.~(\ref{AV1}) are not of comparable sizes. The first is of order $O(\tau)$, whereas $\eta$ is $O(\tau^{1/2})$. This does not mean that $\eta$ is more important, since it tends to cancel in the long run, whereas the first term persists.

The next question concerns the precise value of $x$ at which Eq.~(\ref{AV1}) should be evaluated. Should it be the initial value, $x=x_{\rm in}$, the final value $x=x_{\rm in}+\Delta x$, or some intermediate value? For the first term this question is irrelevant, since the change in the value of this term is $O(\tau\Delta x)$ and its contribution vanishes in the limit $\tau\rightarrow 0$ (even after noting that decreasing the time interval by some factor increases the number of intervals by the same factor). This is not the case for the choice of the $x$-dependent variance of $\eta$.

The It\^{o} procedure evaluates the variance at $x=x_{\rm in}$. $\eta$ has a symmetric distribution and $\langle\eta \rangle=0$. The Stratonovich procedure evaluates the variance at the middle of the interval, $x=x_{\rm in}+\Delta x/2$. In order to distinguish between the two procedures, we denote the respective random terms by $\eta_{\rm I}$ and $\eta_{\rm S}$. In order to fix ideas, let us explore the case that $\Gamma (x)$ is an increasing function of $x$. Since the variance is evaluated at $x\approx x_{\rm in}+\eta_{\rm S}/2$, it means that if $\eta_{\rm S}>0$ (respectively $\eta_{\rm S}<0$), then $\eta_{\rm S}/\eta_{\rm I}>1$ (respectively $\eta_{\rm S}/\eta_{\rm I}<1$). Qualitatively, this means that the distribution of $\eta_{\rm S}$ will have a longer tail than that of $\eta_{\rm I}$ in the positive direction and a shorter tail in the negative direction.

In more quantitative terms, we can establish a one to one correspondence between the values of $\eta_{\rm S}$ and those of $\eta_{\rm I}$, such that
\be
\eta_{\rm S}\approx [\Gamma (x_{\rm in}+\eta_{\rm S}/2)/\Gamma (x_{\rm in})]^{1/2}\eta_{\rm I}
\approx (1+\Gamma '\eta_{\rm S}/2\Gamma)^{1/2}\eta_{\rm I}\approx (1+\Gamma '\eta_{\rm I}/4\Gamma)\eta_{\rm I}
=\eta_{\rm I}+\Gamma '\eta_{\rm I}^2/4\Gamma \; ,
\label{seq}
\ee
where we have written $\Gamma$ as shorthand for $\Gamma (x_{\rm in})$ and $\Gamma'$ for $d\Gamma /dx$ at $x=x_{\rm in}$. The approximations in sequel (\ref{seq}) are $\Delta x\approx \eta_{\rm S}$, expansions to first order in $\Delta x$, and $\eta_{\rm S}\eta_{\rm I}\approx\eta_{\rm I}^2$. The important consequence is that for sufficiently small $\tau$ we have
\be
\langle\eta_{\rm S}\rangle=\langle\eta_{\rm I}\rangle+(\Gamma '/4\Gamma)\langle\eta_{\rm I}^2\rangle=k_BT\Gamma '\tau/2 \;.
\ee
This term is not negligible and does not cancel in the long run.

In a didactic article, van Kampen\cite{Kampen} explains how to translate between the It\^{o} and the Stratonovich procedures. He advocates the use of a master equation rather than a Langevin approach, so that the It\^{o}--Stratonovich dilemma never arises. Lan\c{c}on et al.\ performed an experiment in which colloidal particles diffuse in a medium with position-dependend diffusion coefficient, so that $\langle\eta_{\rm S}\rangle\neq\langle\eta_{\rm I}\rangle$. In their case they found that $\Gamma (x)$ has to be evaluated at $x=x_{\rm in}$.

In order to judge what is the appropriate procedure for Eq.~(\ref{AV1}) in our case, we will evaluate $\langle\eta \rangle$. Since the difference between both procedures depends on $w(x)$ and not on $\lambda $, we are free to take $\lambda\equiv 1$ (we may imagine that $E-k_BT\log\Gamma$ is constant) and are left with a random walk problem in which each step is equally probable for both directions. Let the system be initially at $x=x_{\rm in}$; by definition, the probability distribution for $\eta $ is the probability distribution for $x-x_{\rm in}$ after time $\tau$.

Let $P_i$ be the probability to find the system at $x=i\ell$ at some moment. The change of probability after time $\delta t$ will be $\delta P_i=(w_{i-1}P_{i-1}-2w_{i}P_{i}+w_{i+1}P_{i+1})\delta t$. Or, defining $\rho_i=w_{i}P_{i}$,
\be
\delta \rho _i=w_{i}(\rho_{i-1}-2\rho_{i}+\rho_{i+1})\delta t \;.
\label{discrete}
\ee
In order to use the central limit theorem, we require $\delta t/\tau\rightarrow 0$ and we pass to a continuous model. Equation (\ref{discrete}) becomes the diffusion equation
\be
\frac{\partial \rho}{\partial t}=k_BT\Gamma (x)\frac{\partial^2 \rho}{\partial x^2} \;.
\label{diffusion}
\ee
In order to pass from Eq.~(\ref{discrete}) to Eq.~(\ref{diffusion}) we have expanded $\rho (x)$ to order $O(\ell^2)$ and used the definition of $\Gamma$.
Equation (\ref{diffusion}) may be regarded as our master equation.

Let us take the initial value $\rho(x,0)=w(x_{\rm in})\delta (x-x_{\rm in})$. 
If in Eq.~(\ref{diffusion}) $\Gamma (x)$ were substituted by the constant $\Gamma (x_{\rm in})$, the solution of the diffusion equation would be
\be
\rho^{(0)}(x,t)=\frac{w(x_{\rm in})}{2\sqrt{\pi k_BT\Gamma(x_{\rm in})t}}\exp\left[ -\frac{(x-x_{\rm in})^2}{4k_BT\Gamma(x_{\rm in})t}\right]\;,
\ee
which is just the It\^{o} distribution multiplied by $w(x_{\rm in})$. 
We now deal with Eq.~(\ref{diffusion}) by means of the approximation
\be
\frac{\partial \rho}{\partial t}=k_BT[\Gamma (x_{\rm in})+\Gamma (x)-\Gamma (x_{\rm in})]\frac{\partial^2 \rho}{\partial x^2}
\approx k_BT\Gamma (x_{\rm in})\frac{\partial^2 \rho}{\partial x^2}+k_BT\Gamma '\cdot (x-x_{\rm in})\frac{\partial^2 \rho^{(0)}}{\partial x^2} \;.
\label{approxs}
\ee

\begin{figure}
\scalebox{0.85}{\includegraphics{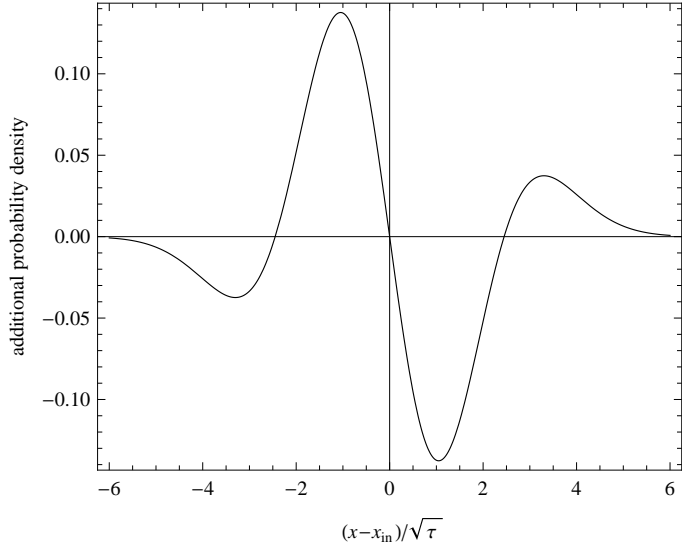}}%
\caption{\label{deviation}Deviation of the probability density from the It\^{o} distribution. The unit of length can be chosen arbitrarily. The unit of time equals the square of the length unit divided by $k_BT\Gamma(x_{\rm in})$. The probability density is in units of $d\log\Gamma/dx$.}
\end{figure}

This nonhomogeneous equation is solved using the Green function of the diffusion equation.\cite{Hand} We obtain that the deviation of the probability distribution from the It\^{o} distribution is 
$\Delta {\cal P}(x,\tau )=(d\log\Gamma/dx)(-6\phi+\phi^3)\exp(-\phi^2/4)/(16\sqrt{\pi})$,
where $\phi=(x-x_{\rm in})/\sqrt{k_BT\Gamma (x_{\rm in})\tau}$ and the derivative is evaluated at $x=x_{\rm in}$. A plot of this deviation is shown in Fig.~\ref{deviation} [for any value of $\tau$ for which the approximations in Eq.~(\ref{approxs}) are justified].

The deviation $\Delta {\cal P}$ enhances the positive tail of the distribution of $\eta_{\rm I}$ and hinders the negative tail, as in the case of $\eta_{\rm S}$. However, the key feature is that $\langle\eta\rangle$ is not affected by $\Delta {\cal P}$; therefore, $\eta_{\rm I}$ has the appropriate distribution for our problem and the It\^{o} procedure should be used. Also $\langle \eta^2\rangle$ is not affected by $\Delta {\cal P}$. It is interesting to note that, no matter how small $\tau$ is, there are always values of $x$ where $\Delta {\cal P}$ remains finite; however, the statistical weight of $\Delta {\cal P}$ does become negligible in the limit of small $\tau$.

\section{Linear transformation of variables\label{transf}}
Let $x_1,\dots x_N$ be the original set of variables for which the evolution of the system is known.
Let us restrict ourselves to cases in which the coefficients $\Gamma _j$ are independent of the coordinates, so that $G_j=E$. Let $u_1,\dots u_N$ be a new set of variables, defined by means of a linear transformation
\be
u_i=\sum_{j=1}^N C_{ij}\frac{x_j}{\sqrt{\Gamma_j}}  \;,
\label{orth}
\ee
where $C_{ij}$ are the elements of a constant orthogonal matrix ${\bf C}$; the purpose of the factor $\sqrt{\Gamma_j}$ is to compensate for the anisotropy of phase space.

Substituting Eq.~(\ref{AVN}) into Eq.~(\ref{orth}) we obtain
\be
\Delta u_i=-\sum_{j=1}^N C_{ij} \sqrt{\Gamma_j}\frac{\partial E}{\partial x_j}\tau +\xi_i \;,
\label{Du}
\ee
where $\xi_i=\sum_{j=1}^N C_{ij}\eta_j/\sqrt{\Gamma_j}$ is a fluctuating term with zero average and Gaussian distribution. Its variance is
\be
\sum_{j=1}^N C_{ij}^2\langle \eta_j^2\rangle /\Gamma_j=2k_BT \tau \sum_{j=1}^N C_{ij}^2=2k_BT \tau \;,
\ee
where we have used the property that $\eta_j$ and $\eta_{j'}$ are not correlated for $j\neq j'$ and orthogonality of ${\bf C}$. For $i\neq i'$ we have
\be
\langle \xi_i\xi_{i'}\rangle=\sum_{j,j'=1}^N C_{ij}C_{i'j'}\langle \eta_j\eta_{j'}\rangle /\sqrt{\Gamma_j\Gamma_{j'}}=2k_BT \tau \sum_{j=1}^N C_{ij}C_{i'j}=0 \;,
\ee
where we have used again $\langle \eta_j\eta_{j'}\rangle=2k_BT \tau \Gamma_j\delta _{jj'}$ and orthogonality of ${\bf C}$.

Since $x_j=\sqrt{\Gamma_j}\sum_{i=1}^N {\bf C}^{-1}_{ji}u_i=\sqrt{\Gamma_j}\sum_{i=1}^N C_{ij}u_i$, $\partial x_j/\partial u_i=\sqrt{\Gamma_j}C_{ij}$. Therefore,
\be
\frac{\partial E}{\partial u_i}=\sum_{j=1}^N \frac{\partial x_j}{\partial u_i}\frac{\partial E}{\partial x_j}=\sum_{j=1}^N \sqrt{\Gamma_j}C_{ij}\frac{\partial E}{\partial x_j} \;.
\label{dGdu}
\ee
Comparing Eq.~(\ref{dGdu}) and Eq.~(\ref{Du}) we finally obtain
\be
\Delta u_i=-\frac{\partial E}{\partial u_i}\tau +\xi_i \;.
\ee

\section{Numeric Test \label{test}}
In this Appendix we evaluate the statistical average of several quantities for a 2D system, using polar coordinates. In the following example we take an energy in which the Cartesian coordinates separate, namely, $E=x^2+2y^2$; the statistical average of any function of $x$ will be $\langle f(x)\rangle=\int_{-\infty}^\infty f(x)\exp[-x^2/(k_BT)]dx/\int_{-\infty}^\infty \exp[-x^2/(k_BT)]dx$ and, similarly, $\langle g(y)\rangle=\int_{-\infty}^\infty g(y)\exp[-2y^2/(k_BT)]dy/\int_{-\infty}^\infty \exp[-2y^2/(k_BT)]dy$ and $\langle f(x)g(y)\rangle=\langle f(x)\rangle\langle g(y)\rangle$. The following is a Mathematica-program that evaluates several averages of this kind.

\begin{table}[b]
\caption{\label{compare}Statistical averages of several quantities, for two temperatures. ``theory" refers to ensemble averages and ``numeric" to average over steps, as obtained from the evolution predicted by Sec.~\ref{polar}.}
\begin{tabular}{l|cccccc}
& $\langle xy\rangle$ & $\langle |x|\rangle$ & $\langle x^2 \rangle$ & $\langle |y|\rangle$ & $\langle y^2\rangle$ & $\langle |xy|\rangle$ \\
\hline 
theory, $k_BT=1$ & ~0.00 & 0.564 & 0.500 & 0.399 & 0.250 & 0.225 \\
numeric, $k_BT=1$ & -0.01 & 0.561 & 0.493 & 0.402 & 0.253 & 0.227 \\
theory, $k_BT=2$ & ~0.00 & 0.798 & 1.000 & 0.564 & 0.500 & 0.450 \\
numeric, $k_BT=2$ & -0.02 & 0.813 & 1.028 & 0.557 & 0.490 & 0.456
\end{tabular}
\end{table}

\begin{verbatim}\end{verbatim}
\ni\verb#Clear[r, phi];   (*# These are the polar coordinates $r,\varphi $ \sf
\ni\verb#energy = r^2 (1 + Sin[phi]^2);   (*# This is the energy in polar coordinates \sf
\ni\verb#kT = 2;   (*# This is $k_BT$ \sf
\ni\verb#Gammartau = 5.*^-4 r; Gammaphitau = 5.*^-5/r^2;   (*# These are $\Gamma_r\tau$ and $\Gamma_\varphi \tau$; only the products of these quantities appear in each step; statistical averages should be independent of these dynamical functions, which have been chosen arbitrarily \sf
\ni\verb#stepr = Simplify[Gammartau D[kT Log[r Gammartau] - energy, r]];   (*# This is $\Delta r$ as in Eq.~(\ref{AV1pol}) without $\eta_r$  \sf
\ni\verb#stepphi = Simplify[Gammaphitau D[kT Log[r Gammaphitau] - energy, phi]];   (*# This is $\Delta \varphi$ without $\eta_\varphi $  \sf
\ni\verb#stdr = Simplify[Sqrt[2 kT Gammartau], Assumptions -> r > 0]; stdphi =# 
\ni\verb# Simplify[Sqrt[2 kT Gammaphitau], Assumptions -> r > 0];   (*# These are the standard deviations of $\eta_r$ and $\eta_\varphi $  \sf
\ni\verb#r = 0.1; phi = 0.2;   (*# This is the initial microstate; the statistical averages should be independent of it  \sf
\ni\verb#Nrelax = 5 10^6;   (*# Number of steps during which the system ``forgets" the initial state and relaxes from it to a ``typical" microstate \sf
\ni\verb#Do[r = r + stepr + RandomReal[NormalDistribution[0, stdr]];   (*# $r$ evolves during a step, according to Eq.~(\ref{AV1pol}). \verb#stdr# is evaluated at the beginning of the step, according to the It\^{o} procedure \sf 
\ni\verb#  If[r < 0, r = -r; phi = phi - Pi];   (*# If $r$ becomes negative, $r$ and $\varphi$ are redefined \sf 
\ni\verb#  phi = phi + stepphi + RandomReal[NormalDistribution[0, stdphi]],   (*# $\varphi $ evolves during a step, according to Sec.~\ref{polar}. $\Gamma_\varphi$ does not depend on $\varphi$, so that the stage at which \verb#stdphi# is evaluated is not crucial \sf
\ni\verb#     {i,Nrelax}];   (*# Length of the loop \sf
\ni\verb#Naverage = 15 10^6;   (*# Number of steps during which averages will be evaluated \sf
\ni\verb#sx = 0; sxx = 0; sy = 0; syy = 0; sxy = 0; sabs = 0;   (*# Initialization of the variables that will be used for the evaluation of cumulative sums, from which the averages will be obtained; the following loop is identical to the one above, except that now we keep track of these sums \sf
\ni\verb#Do[r = r + stepr + RandomReal[NormalDistribution[0, stdr]];# 
\ni\verb#  If[r < 0, r = -r; phi = phi - Pi];#\newline 
\ni\verb#  phi = phi + stepphi + RandomReal[NormalDistribution[0, stdphi]];# 
\ni\verb#  x = r Cos[phi]; y = r Sin[phi]; sxy = sxy + x y; x = Abs[x];# 
\ni\verb#  y = Abs[y]; sx = sx + x; sxx = sxx + x^2; sy = sy + y;# 
\ni\verb#  syy = syy + y^2; sabs = sabs + x y, {i, Naverage}];#
\ni\verb#Print["<xy>=", sxy/Naverage]; Print["<|x|>=",sx/Naverage]; #\newline 
\ni\verb# Print["<x^2>=", sxx/Naverage]; Print["<|y|>=",sy/Naverage]; #\newline 
\ni\verb# Print["<y^2>=", syy/Naverage]; Print["<|xy|>=",sabs/Naverage]; (*# The averages that we decided to evaluate are printed \sf 

In Table~\ref{compare} we compare the results obtained by this program with the expected statistical averages.


\begin{thebibliography}{99}
\bibitem{Lang} P. Langevin, ``Sur la th\'{e}orie du mouvement brownien," C. R. Acad. Sci. (Paris) {\bf 146}, 530--533 (1908). See also D. S. Lemons and A. Gythiel, ``Paul Langevin's 1908 paper On the Theory of Brownian Motion," Am. J. Phys. {\bf 65}, 1079--1081 (1997).
\bibitem{Gi}D. Gillespie, ``Fluctuation and dissipation in Brownian motion," Am. J. Phys. {\bf 61}, 1077--1083 (1993).
\bibitem{KT}Y. Katayama and R. Terauti, ``Brownian motion of a single particle under shear flow," Eur. J. Phys. {\bf 17}, 136--140 (1996).
\bibitem{Bal}R. Balescu, ``Stochastic transport in plasmas," Eur. J. Phys. {\bf 21}, 279--288 (2000).
\bibitem{BrinI}E. Bringuier, ``On the Langevin approach to particle transport," Eur. J. Phys. {\bf 27}, 373--382 (2006). 
\bibitem{BrinII}E. Bringuier, ``From mechanical motion to Brownian motion, thermodynamics and particle transport theory," Eur. J. Phys. {\bf 29}, 1243--1262 (2008). 
\bibitem{Chand}S. Chandrasekhar, ``Stochastic problems in physics and astronomy," Rev. Mod. Phys. {\bf 15}, 1--89 (1943).
\bibitem{HH} P. C. Hohenberg and B. I. Halperin, ``Theory of dynamic critical phenomena," Rev. Mod. Phys. {\bf 49}, 435--479 (1977).
\bibitem{Cof} W.T. Coffey, Yu.P. Kalmykov and J.T. Waldron, {\it The Langevin Equation: with Applications to Stochastic Problems in Physics, Chemistry, and Electrical Engineering} 2nd ed. (World Scientific, Singapore, 2004).
\bibitem{relat}J. Dunkel and P. H\"{a}nggi, ``Relativistic Brownian motion," Phys. Rep. {\bf 471}, 1--73 (2009).
\bibitem{FDT1}H. B. Callen and R. F. Greene, ``On a theorem of irreversible thermodynamics," Phys. Rev. {\bf 86}, 702--710 (1952).
\bibitem{FDT2}F. Reif, {\it Fundamentals of Statistical and Thermal Physics} (McGraw-Hill, Kogakusha, 1965).
\bibitem{FDT3} R. Kubo, M. Toda, and N. Hashitsume, {\it Statistical Physics II} (Springer, Berlin, 1995).
\bibitem{Str}R. L. Stratonovich, {\it Conditional Markov Processes and Their Application to the Theory of Optimal Control} (Elsevier, New York,1968).
\bibitem{Schu} Z. Schuss, {\it Theory and Applications of Stochastic Differential Equations} (Wiley, New York, 1980).
\bibitem{Oks}B. K. {\O}ksendal, {\it Stochastic Differential Equations} 6th ed. (Springer, 2003). 
\bibitem{CL} A. I. Khinchin, {\it Mathematical Foundations of Statistical Mechanics} (Dover, New York, 1949) p. 166.
\bibitem{Fel}W. Feller, {\it An Introduction to Probability Theory and its Applications}, 2nd ed., Vol. II (Willey, New York, 1971).
\bibitem{Engel}M. Raible and A. Engel, ``Langevin equation for the rotation of a magnetic particle," Appl. Organometal. Chem. {\bf 18}, 536--541(2004).\bibitem{GGG}H. Grabert and S.M. Green, ``Fluctuations and nonlinear irreversible processes," Phys. Rev. A {\bf 19}, 1747--1756 (1979), H. Grabert, R. Graham and S.M. Green, ``Fluctuations and nonlinear irreversible processes II," Phys. Rev. A {\bf 21}, 2136--2146 (1980); for a readable summary see P. H\"{a}nggi, ``Connection between deterministic and stochastic descriptions of nonlinear systems," Helv. Phys. Acta {\bf 53}, 491--496 (1980).
\bibitem{GL}L. P. Gor'kov and G. M. Eliashberg, Zh. Eksp. Teor. Fiz. {\bf 54}, 612--626 (1968) [``Generalization of the Ginzburg-Landau equations for non-stationary problems in the case of alloys with paramagnetic impurities," Soviet Phys. JETP {\bf 27}, 328--334 (1968)];
A. Schmid, ``A time dependent Ginzburg--Landau equation and its application to the problem of resistivity in the mixed state," Phys. Kondens. Mater. {\bf 5}, 302--317 (1966).
\bibitem{Tinkham} M. Tinkham, {\it Introduction to Superconductivity} (Dover, 1996); N. B. Kopnin, {\it Theory of Nonequilibrium Superconductivity} (Oxford University, 2001).
\bibitem{Kramer}L. Kramer and R. J. Watts-Tobin, ``Theory of dissipative current-carrying states in superconducting filaments," Phys. Rev. Lett. {\bf 40}, 1041--1044 (1978); R. J. Watts-Tobin, Y. Kr\"{a}henb\"{u}hl, and L. Kramer, ``Nonequilibrium theory of dirty, current-carrying superconductors: phase-slip oscillators in narrow filaments near $T_c$,"  J. Low Temp. Phys. {\bf 42}, 459--501 (1981).
\bibitem{Kampen}N. G. van Kampen, ``It\^{o} versus Stratonovich," J. Stat. Phys. {\bf 24}, 175--187 (1981).
\bibitem{Lancon}P. Lan\c{c}on, G. Batrouni, L. Lobry, and N. Ostrowsky, ``Brownian walker in a confined geometry leading to a space-dependent diffusion coefficient," Physica A {\bf 304}, 65--76 (2002). 
\bibitem{Hand}A. D. Polyanin, {\it Handbook of Linear Partial Differential Equations for Engineers and Scientists} (Chapman \& Hall/CRC, 2002).
\end{thebibliography}
\end{document}